\DocumentMetadata{
	lang		= eng-US, 	% default
	pdfstandard	= A-3u,		% A-2b, A-2u, A-3b, A-3u. Check that your figures also meet the standard you choose.
	pdfversion	= 1.7, 		% an older version of PDF, but very widely supported
}

\documentclass[colorlinks,upint,subscriptcorrection,varvw,hyphenate,balance,greek,russian,vietnamese,german]{asmeconf} % pdftex
\hypersetup{%
	pdfauthor={Chaitanya S Nayak},									  % <=== change to YOUR name
	pdftitle={FEDSM-ASME Manuscript},                  % <=== change to YOUR pdf file title
	pdfkeywords={Machine Learning, Framework, Bubbles},% <=== change to YOUR pdf keywords
	pdfsubject = {ML on Fluids},			  % <=== change to YOUR subject
}
\allowdisplaybreaks % from amsmath package, allows multiline equations to break across pages (delete if not wanted)
\begin{document}

% Change these fields to the right content for your conference.
% You can comment these out if for some reason you don't want a header.
% Use title case for the conference name (first letters capitalized), not all capitals

\ConfName{Proceedings of the ASME 2026\linebreak Fluids Engineering Division Summer Meeting}
\ConfAcronym{FEDSM2026}
\ConfDate{July 26--29, 2026} 
\ConfCity{Bellevue, WA}
\PaperNo{FEDSM2026-183961}

% Units of measure (e.g., cm) and other specialty lowercase terms in the title should be 
%   enclosed in \NoCaseChange{...} to maintain lower case type.
%   The rest of the title will automatically be set in all capital letters.
%
%	\title{Place Title Here: Place Subtitle After Colon} 

\title{Depth-Aware Machine Learning Framework for Bubble Characterization in Two-Phase Flows} % <=== replace with YOUR title
 
%   Put author names into the order you want. Use the same order for affiliations.
%   \affil{#} tags the author's affiliation to the address in \SetAffiliation{#}.
%   No space between last name and \affil{#}, separate names with commas.
%
%	For a sole author or a single affiliation for all authors, {#} may be left empty, i.e. \affil{} and \SetAffiliation{} (but not with [grid] option!)
%
%   \CorrespondingAuthor{email} follows that author's affiliation, no spaces before or after 
%   If multiple corresponding authors, put both email addresses in the same command and place after both authors.
%
%   \JointFirstAuthor, if applicable, follows the affiliation of the relevant authors, no spaces.

\SetAuthors{%
    Chaitanya S Nayak\affil{1}\affil{2}\CorrespondingAuthor{}\JointFirstAuthor{},
    Faizaan Mohammed\affil{1}\affil{2}\JointFirstAuthor{},
    Vivek Kumar\affil{1}\affil{2},
    Shivam Prajapati\affil{1}\affil{2},
    Cyrus Aidun\affil{1}\affil{2}\CorrespondingAuthor{cyrus.aidun@me.gatech.edu, cnayak6@gatech.edu}
}

\SetAffiliation{1}{George W Woodruff School of Mechanical Engineering, Georgia Institute of Technology, Atlanta, Georgia}
\SetAffiliation{2}{Renewable Bioproducts Institute, Atlanta, Georgia}

%   Note: You can force a line break in the address using \\ 

%	To switch from inline author names to gridded names, use the [grid] option.

\maketitle

%%% Use this footnote for tracking various versions of your draft. Change text to suit your own needs. 
%%% \date{..} calls the same command. 
%%\versionfootnote{Documentation for \texttt{asmeconf.cls}: Version~\versionno, \today.}% <=== Delete before final submission.

%%% Change the following to your keywords.  Keywords are automatically printed at the end of the abstract.
%%% This command MUST COME BEFORE the end of the abstract.
%%% If you don't want keywords, leave the argument of \keywords{} empty (or use the abstract* environment)


%%%%%  End of fields to be completed. Now write your paper. %%%%%%%%%%%%%%%%%%%%%%%%%%%%%%%%%%%%%%%%%%%

%%%%%  ABSTRACT  %%%%%%%%%%%%%%%%%%%%%%%%%%%%%%%%%%%%%%%%%%%%%%%%%%%
%%
%% Abstract should be 200 words or less

\begin{abstract}
Understanding the three-dimensional motion of bubbles is essential for interpreting transport and mixing in multiphase flows, especially when bubbles deform under shear or move rapidly through the flow field. In many laboratory setups, only a single high-speed camera is available, which limits measurements to two dimensions. Traditional image-processing tools can identify bubbles only when they appear circular and isolated, but they struggle with irregularly shaped bubbles, shear-induced deformations, strong blurring, and partial overlaps. Multi-camera systems could overcome these issues, but require significant hardware additions and calibration effort.

In this work, we introduce a new machine-learning framework that can detect bubbles and estimate their depth using only a single 20 kHz high-speed camera with 3 \textmu m resolution. The method first uses a large unlabeled dataset and clusters the bubbles with an unsupervised algorithm to reveal their underlying structure. These clusters provide pseudo labels, which are combined with a small set of true in-plane bubble labels to train a semi-supervised model that generalizes across different bubble appearances. These components produce a continuous depth-proxy score that indicates how close each bubble is to the imaging plane, even when bubbles are distorted or irregularly shaped. In parallel, we perform robust bubble identification using instance segmentation, which separates touching, overlapping, and elongated bubbles generated by high-velocity shear. Quantitatively, the in-plane segmentation baseline achieves strong held-out performance with Average Precision (AP) = 0.818, implying stable detection across thresholds, clutter, bubble detection Precision of 0.901, and a False-Positive Rate (FPR) near 6.1\%, hence low spurious bubbles and cleaner statistics under the tested acquisition conditions.

Overall, this framework provides a low-cost, hardware-free way to detect irregularly shaped bubbles and estimate their depth, and temporally track their motion. It can be directly extended to other multiphase systems involving droplets, particles, capsules, fibers and deformable interfaces.

% The resulting depth-aware trajectories enable insights not possible with standard 2D analysis. Clustering the trajectories reveals distinct dynamical regimes, depth-segregated migration paths, and shear-driven dispersion patterns that would normally require multi-camera setups to observe. Because the approach relies on learned image features rather than geometric assumptions, it remains robust at higher void fractions, large velocity gradients, and severe bubble deformation. 

\end{abstract}

%%%%%%%%%  NOMENCLATURE (OPTIONAL) %%%%%%%%%%%%%%%%%%%%%%%%%%%%%%%%%
%%
%% To change space between the symbols and  definitions, use \begin{nomenclature}[Xcm] where X is a number 
%% The unit cm can be replaced by any LaTeX unit of dimension: pt, in, ex, em, pc, etc.
%% Default is 2em.
%%
%% \EntryHeading{..} produces an italicized subheading in the nomenclature list, e.g., \EntryHeading{Greek letters}

%%%%%%%%%  BODY OF PAPER %%%%%%%%%%%%%%%%%%%%%%%%%%%%%%%%%

\section{Introduction}
Quantitative bubble or particle tracking underpins a wide class of multiphase-flow diagnostics because it enables time-resolved measurement of transport processes such as mass, momentum, and energy through the motion and evolution of dispersed entities. In many practical systems, these macroscopic transport outcomes are controlled by particle-scale dynamics, including size distributions, relative velocities, accelerations, deformation, and interaction events such as collision, coalescence, and breakup. For instance, in biomedical applications, tracking the shape of deformed Taylor bubbles during the separation of red blood cells from plasma is critical~\cite{kumar2023particle}. For dynamic surface tension measurements using the pendant drop method, the droplet contour must be accurately fitted and its edges precisely detected to evaluate the shape factor and subsequently determine the surface tension~\cite{kumar2025viscosity}. For this edge detection task, Sena et al. (2025) employed a machine-learning-based algorithm; however, their approach cannot reliably distinguish between in-focus and out-of-focus droplets~\cite{sena2025machine}. Therefore, robust identification of the focused droplet remains essential for accurate surface tension evaluation. Robustly identifying and tracking individual particles across image sequences is therefore a prerequisite for extracting physically meaningful statistics and for developing predictive models of complex flows. This work focuses on bubbly gas–liquid flows, where dispersed bubbles interact strongly with the carrier liquid and can significantly alter turbulence, momentum exchange, and interfacial transport.

Optical analysis of bubbly flows remains challenging because bubbles are non-rigid and transient objects: they deform continuously, overlap in projection, merge or fragment, and may intermittently enter or exit the imaging plane. These effects complicate both reliable instantaneous interface tracking and temporal association of bubbles across frames. Yet resolving these dynamics is crucial, since bubble size distributions and interfacial area density govern heat and mass transfer, mixing efficiency, and overall flow stability.

Classical image-processing pipelines for particle detection typically rely on filtering and thresholding heuristics. For instance, Laplacian-of-Gaussian filtering provides a principled scale-selective operator for blob-like feature enhancement and has been widely used as a foundation for feature detection in noisy images \cite{sotak1989_log}. Dimensionality-reduction methods such as principal component analysis (PCA) have also been used to concentrate dominant variance modes and suppress noise/background structure in high-dimensional signals and image-derived features \cite{mackiewicz1993_pca}. However, in bubbly flows these approaches can become brittle: illumination gradients, specular highlights, strong overlap, and continuous shape change can drive failure modes and increase manual tuning effort. Critically, many traditional pipelines yield only coarse geometric descriptors rather than pixel-resolved interfaces, limiting downstream analysis of interfacial geometry.

Recent learning-based approaches have improved robustness by shifting from handcrafted features to data-driven detection and segmentation. Haas et al. introduced BubCNN, combining a Faster R-CNN detector with a dedicated shape regression network for bubble characterization, demonstrating strong performance across varying conditions \cite{haas2020_bubcnn}. Toyama et al. applied Faster R-CNN-based detection with tracking to quantify hydrogen and oxygen bubbles in alkaline water electrolysis imaging, illustrating the viability of deep detection + tracking pipelines in visually challenging bubble datasets \cite{toyama2025_fasterrcnn}. Going beyond bounding boxes, Kim and Park trained a learning-based model for automated bubble mask extraction in complex gas-liquid two-phase flows, enabling pixel-level interfaces suitable for geometric and interfacial analyses \cite{kim2021_universal}. More broadly, segmentation performance and boundary fidelity depend strongly on architectural choices (e.g., encoder–decoder vs. instance-segmentation frameworks) and loss-function design, motivating careful selection of training objectives for interface-sensitive prediction \cite{azad2025_losses_segmentation,gfg_unet_2025,sinha2025_maskrcnn_unet}.

Despite recent advances, a key limitation of most bubble imaging methods is that their outputs are still effectively two-dimensional, even though the relevant physics and statistics are three-dimensional. Estimating depth usually requires multiple cameras, along with complex calibration, higher costs, and larger equipment footprints, which are often impractical in many lab and industrial environments. To address this, the present work develops a single-camera, depth-aware bubble identification and tracking method that maintains accurate bubble interfaces while remaining practical for constrained experimental setups.

Gaussian mixture modeling (GMM) has been demonstrated as an effective unsupervised clustering mechanism for semantic segmentation in challenging imaging settings \cite{kartakoullis2025_gmm}, and standard model selection strategies provide a systematic route to tune such models \cite{sklearn_gmm_selection_2025}. Similarly, novelty and outlier detection formulations offer a structured way to flag observations that deviate from the learned distribution of valid objects, supporting robustness in heterogeneous image sequences \cite{sklearn_outlier_2025}. Related work combining deep learning with classical machine learning for classification further motivates hybrid pipelines in settings where generalization and interpretability are both valued \cite{xi2022_rf_dl}.

Although bubbly flows motivate the present study, the underlying challenge is tracking deformable, intermittently occluded objects in heterogeneous imaging conditions which appears broadly across particle-laden systems. Examples include emulsification in food processing (droplet statistics tied to rheology), fiber suspensions in paper and pulp manufacturing (structure formation and mechanical properties), and other industrial multiphase processes where reliable size and interface estimates are essential. Multiphase research programs targeting industrially relevant forming and transport processes further emphasize the need for scalable, experimentally practical diagnostics for dispersed flows \cite{gatech_rbi_multiphase_2025}.

Recent machine learning approaches have also improved bubble detection and tracking in complex multiphase flows by enabling robust, pixel-level identification of deformable bubbles \cite{Choi2022BubbleVelocimetry}. Extensions of these methods toward three-dimensional detection and tracking using deep learning further demonstrate the potential for depth-resolved bubble characterization in high void-fraction regimes \cite{Hessenkemper2024_3DBubbleTracking}. However, the reliance on large labeled datasets and computationally intensive models motivates alternative depth-aware frameworks that leverage physically interpretable features and minimal supervision.

Accordingly, the objective of this work is to develop a generalizable bubble identification and tracking pipeline that primarily extracts accurate pixel-resolved bubble interfaces while maintaining robust temporal association under deformation and overlap and finally introduces depth awareness from single-camera sequences to improve size distribution and interfacial area estimates without requiring multi-view imaging.

Motivated by these constraints, we adopt a two-stage strategy that first leverages unsupervised structure in image-derived features to obtain robust bubble candidates and focus-aware groupings and then uses supervised learning to refine pixel-level interfaces and improve generalization across operating conditions.

\section{Dataset Generation}

The images analyzed in this study were acquired using high-speed shadowgraphy in a developing, decaying turbulent bubbly flow established within a square duct. The experiments were performed in a closed-loop facility in which turbulence was generated by a regenerative multiphase pump, producing an initially intense but spatially decaying turbulent field downstream of the pump outlet. A Photron FASTCAM Nova R2 camera coupled with a high-magnification Navitar lens and uniform halogen backlighting was employed to record bubble shadows at high spatial and temporal resolution. Image sequences were captured at multiple axial and radial locations along the duct to resolve the evolution of the bubble population as the turbulence decayed. The raw frames were pre-processed through background subtraction and contrast enhancement, followed by edge detection and watershed segmentation to identify individual bubbles and compute their projected areas. Equivalent diameters were then determined assuming axisymmetric projection, and statistical convergence was ensured by analyzing more than 2000 bubbles for each operating condition. Further details of the experimental configuration, imaging arrangement, and post-processing methodology are provided in Kumar et al. (2025) \cite{kumar2025bubble}, while the characteristics of the pump-generated turbulence are described in Javadi et al (2026)~\cite{javadi2026large}.

%%%%%%%%%%%%%%%%%%%%%%%%%%%%%%%%%%%%%%%%%%%%%%%%%%%%%%%%%%%

\section{METHODS \& RESULTS - UNSUPERVISED LEARNING}

\subsection{Overview of the GMM Computational Pipeline}

This section describes the unsupervised component of the pipeline. It identifies candidate bubble instances and embeds them in a feature space that separates in-focus from out-of-focus populations. Temporal consistency is enforced across successive frames.
Beyond instance detection, the unsupervised stage also enables a depth-aware, pseudo-three-dimensional visualization. It assigns each bubble a relative depth score derived from physically motivated image features. This score provides an estimate of the bubble’s position along the optical axis. 

Classical edge-detection methods such as Canny filtering are robust and effective for delineating bubble boundaries. However, our approach uses edge information as part of a broader feature representation that encodes depth-related cues. Edges are therefore not treated as an isolated geometric output. The resulting unsupervised predictions not only yield interpretable depth-aware visualizations but also serve as structured pseudo-labels for subsequent supervised learning. 
% When combined with supervised refinement, this hybrid strategy forms a generalized and physically informed framework for analyzing complex multiphase systems, enabling robust extraction of interfacial and depth-dependent information from single-camera measurements.

Within this unsupervised framework, GMM clustering is employed to partition bubbles in the learned feature space, enabling separation based on combined sharpness, scale, and depth-related metrics. To promote physically consistent labeling over time, a soft temporal stabilization strategy is applied, which uses inter-frame proximity and feature continuity to suppress spurious label switching while preserving genuine bubble dynamics. The pipeline operates on individual frames without requiring labeled training data and is designed to remain robust to illumination gradients, partial bubbles at image boundaries, and modest inter-frame motion, making it suitable for high-speed imaging conditions and experimentally constrained optical setups.

% \begin{figure}[htpb]
% \centering
% \includegraphics[
%     width=\columnwidth,
%     keepaspectratio
% ]{depth-stack.pdf}
% \caption{Frame visualization along the z-coordinate represents a dimensionless depth proxy computed from the radial distribution of edge sharpness; it is not a calibrated physical depth measurement.}
% \label{fig:depth-stack}
% \end{figure}

% The pseudo-three-dimensional stacked-plane visualization shown in Fig.~\ref{fig:depth-stack} illustrates the depth-aware organization produced by the unsupervised stage on a frame-by-frame basis. This representation does not constitute a geometric depth reconstruction; rather, translucent planes correspond to level sets of the learned depth proxy derived from rim-edge sharpness and intensity-gradient features used in the GMM clustering. Bubbles associated with lower-index planes (green) exhibit optically sharp rims indicative of near-focus regions, whereas those mapped to higher-index planes (red) display progressively diffuse boundaries consistent with increasing defocus. By preserving in-plane spatial relationships while encoding relative focal ordering, this visualization provides an interpretable depiction of how image-derived features are mapped to physically meaningful depth layers.

\begin{figure}[htpb]
\centering
\includegraphics[
    width=\columnwidth,
    keepaspectratio
]{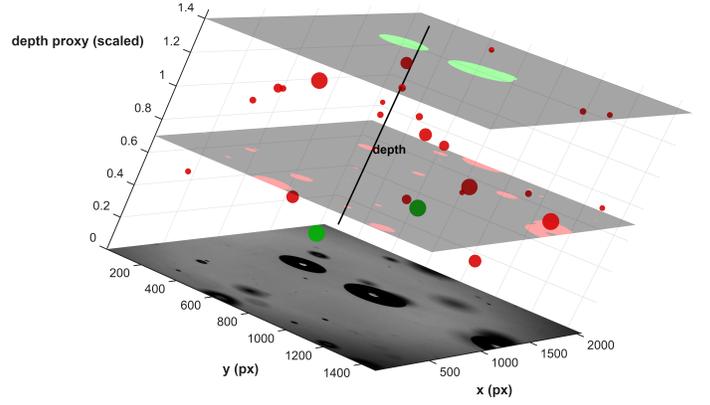}
\caption{Regenerated bubble clouds for pseudo-3D visualization across depth for one frame}
\label{fig:planar-thresholding}
\end{figure}

Figure~\ref{fig:planar-thresholding} provides a pseudo-three-dimensional visualization that links the unsupervised clustering directly to an interpretable optical cue, namely rim-edge sharpness. For a representative frame, the grayscale image is displayed as a reference plane, while semi-transparent overlays indicate the spatial support of bubbles classified as in-focus (green) and out-of-focus (red). Each bubble is positioned at its in-plane centroid and assigned a vertical coordinate corresponding to a normalized depth proxy computed from the radial distribution of intensity gradients along the bubble rim. The resulting separation of focal populations along the depth axis illustrates how the GMM organizes bubbles into physically meaningful layers based on edge-informed features while retaining the spatial structure of the underlying flow.

\subsection{Image Preprocessing and Illumination Correction}\label{sec:illuminationcorrection}

Each frame was first converted to grayscale. A static background image $B(x,y)$ was then constructed to capture large-scale illumination structure while suppressing bubble-scale features. This background was generated by applying a Gaussian blur with a very large standard deviation $\sigma_x = 300$ pixels to a representative raw frame, ensuring that only slowly varying intensity components were retained. This same background image was reused for all frames to reduce computational overhead and to avoid introducing frame-dependent residual patterns that could otherwise be exploited by the model. Illumination conditions were maintained stationary throughout the experiment to prevent the introduction of learnable background bias.

Each frame was then corrected using multiplicative flat-field normalization to compensate for spatially nonuniform illumination. A global scaling factor $s$ was first computed as the mean intensity of the background image,
\begin{equation}
s = \mathrm{mean}(B),
\label{eq:background_mean}
\end{equation}
and the corrected image intensity $I_{\mathrm{corrected}}(x,y)$ was then obtained as
\begin{equation}
I_{\mathrm{corrected}}(x,y) = \frac{s \, I(x,y)}{B(x,y) + \epsilon},
\label{eq:flatfield_correction}
\end{equation}
where $I(x,y)$ denotes the raw image intensity and $\epsilon$ is a small stabilizing constant introduced to prevent numerical amplification in dark regions. In this study, $\epsilon$ was set to
\begin{equation}
\epsilon \approx 10^{-3}.
\label{eq:epsilon}
\end{equation}
This normalization ensures that the corrected image remains within a comparable intensity range across frames, improving the consistency of intensity- and gradient-based features and facilitating reliable transfer of thresholds and learned decision boundaries.

\subsection{Bubble Segmentation}

Following illumination correction, Binary segmentation was performed using Otsu’s method, \cite{Otsu1979}. Bubble candidates are identified from the contrast-enhanced image by applying this thresholding method that exploits the bimodal separation between bubble interiors and background regions. This step converts the continuous-valued image into an initial binary representation that serves as the basis for subsequent morphological refinement and instance extraction.

Specifically, a binary mask $B(x,y)$ is obtained, which selects an optimal threshold $T_{\mathrm{Otsu}}$ by maximizing the inter-class variance of the grayscale intensity distribution. The resulting segmentation is defined as
\begin{equation}
B(x,y) =
\begin{cases}
1, & I_c(x,y) < T_{\mathrm{Otsu}}, \\
0, & \text{otherwise},
\end{cases}
\label{eq:otsu}
\end{equation}
where $I_c(x,y)$ denotes the illumination-corrected image intensity at pixel location $(x,y)$. Pixels assigned $B(x,y)=1$ correspond to candidate bubble regions.

The initial binary mask is further refined using a sequence of morphological operations, including opening and closing to remove small-scale noise and bridge fragmented regions, hole filling to recover bubble interiors, and small-object removal to suppress isolated artifacts. These operations enforce spatial coherence while preserving bubble-scale structures.

Connected components are then extracted from the refined binary mask and filtered using physically motivated geometric constraints. Components are retained only if their area lies within prescribed minimum and maximum thresholds, their solidity and circularity exceed specified limits, and their bounding-box aspect ratio remains within an admissible range. This filtering strategy suppresses elongated background structures and large dark regions associated with image boundaries or floor artifacts, while retaining partially visible bubbles near the edges of the field of view.

% Each retained connected component is treated as an individual bubble instance, and its corresponding binary mask is propagated to subsequent stages for feature extraction and depth-aware clustering.

%%%%%%%%%%%%% begin figure %%%%%%%%%%%%%%%%%

%% captions go below figures

\begin{figure*}[t]
\centering
\includegraphics[width=\textwidth,height=0.48\textheight,keepaspectratio]{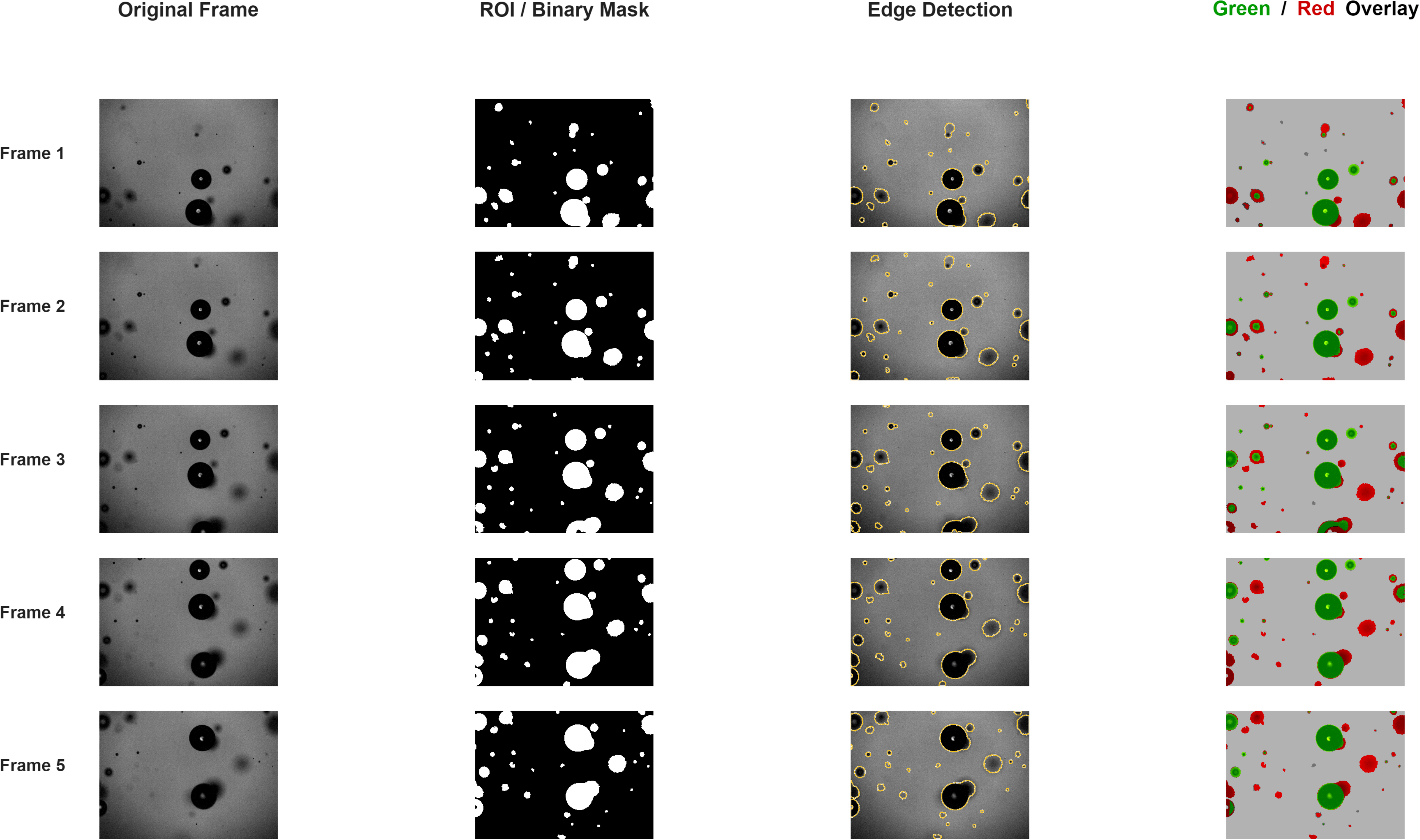}
\caption{Depth-aware bubble segmentation, overlap detection and focal-plane clustering across five consecutive frames; In-Focus: Green Clusters, Out-of-Focus: Red Clusters}
\label{fig:low-vf-clustering}
\end{figure*}

%%%%%%%%%%%%% end figure %%%%%%%%%%%%%%%%%%%

%%%%%%%%%%%%% begin figure %%%%%%%%%%%%%%%%%

%% captions go below figures

\begin{figure*}[t]
  \centering
  \includegraphics[width=0.95\textwidth]{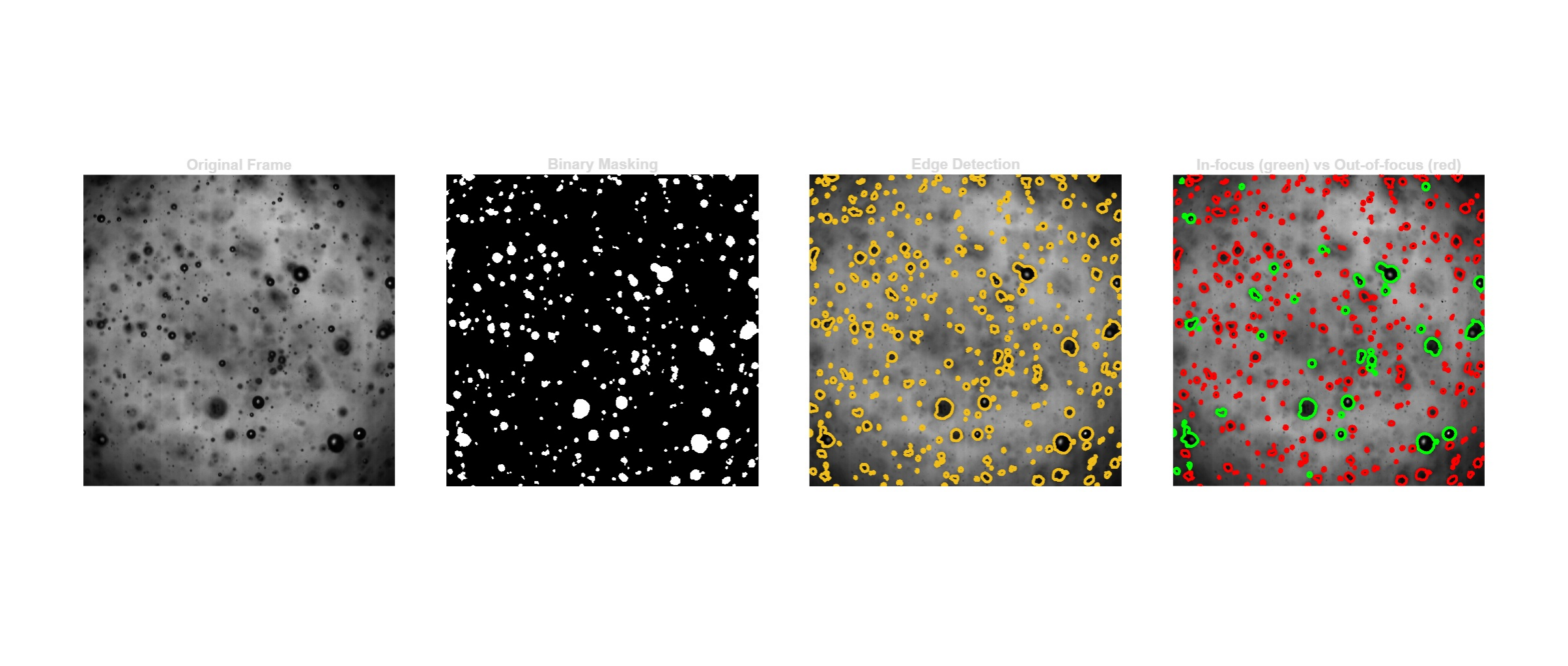}
  \caption{Bubble segmentation and clustering for a higher void fraction}
  \label{fig:high-vf-clustering}
\end{figure*}

%%%%%%%%%%%%% end figure %%%%%%%%%%%%%%%%%%%

\subsection{Engineered Feature Extraction}\label{sec:ULfeature}

With individual bubble instances delineated through segmentation, a set of engineered features was extracted to encode optical cues related to focal depth while remaining invariant to absolute bubble size. For each segmented bubble, a localized image crop centered at the bubble centroid was extracted and rescaled to a fixed spatial resolution, ensuring consistent feature computation across bubbles of varying physical size.

To capture focus-dependent edge information, a focus-sensitive metric referred to as radial sharpness was computed from the cropped intensity field. Image gradients were evaluated within concentric annular regions extending inward from the bubble boundary, with greater emphasis placed on gradients near the rim. This weighting reflects the physical observation that bubbles located near the focal plane exhibit sharper intensity transitions at the gas--liquid interface due to stronger refractive-index contrast, whereas defocused bubbles display more diffused edge gradients.

The local gradient magnitude at pixel location $(x,y)$ was computed from the illumination-corrected image $I_c(x,y)$ as
\begin{equation}
\nabla I_c(x,y) =
\sqrt{
\left(\frac{\partial I_c}{\partial x}\right)^2 +
\left(\frac{\partial I_c}{\partial y}\right)^2
},
\label{eq:grad}
\end{equation}
where the partial derivatives quantify spatial intensity variations along the horizontal and vertical directions. Radial sharpness was then obtained as a weighted aggregation of gradient magnitudes across successive annuli, with weights inversely proportional to the radial distance from the bubble boundary, thereby prioritizing edge-localized information.

In addition to edge sharpness, a radial depth proxy was defined to characterize the spatial distribution of gradient energy within the bubble interior. This metric was computed as the gradient-weighted mean radial distance measured inward from the bubble boundary, providing a scalar indicator of whether gradient strength was concentrated near the rim or distributed deeper within the bubble. Bubbles exhibiting sharp but radially diffuse gradients were therefore associated with out-of-plane or defocused structures, whereas bubbles with gradients strongly localized near the boundary were indicative of in-plane, well-focused bubbles.

Bubble area and centroid position were also recorded for each detected instance. These quantities support physical consistency checks, such as identifying dominant foreground bubbles, and enable association of bubbles across consecutive frames during temporal stabilization. 
% Collectively, these physically informed features provide a robust representation for subsequent unsupervised clustering, enabling reliable discrimination between in-focus and out-of-focus bubbles under varying illumination conditions and partial occlusion.

\subsection{Clustering}

Building on the engineered feature representation described above, clustering was performed to separate bubbles into distinct focal populations. For each detected bubble, a feature vector $\tilde{\mathbf{x}} \in \mathbb{R}^3$ was constructed from the radial sharpness metric, the radial depth proxy, and the logarithm of the bubble area. Prior to clustering, each feature dimension was normalized to unit variance to prevent dominance by any single feature and to ensure balanced contribution within the clustering process.

The distribution of feature vectors was modeled using a GMM with two components, corresponding to in-focus and out-of-focus bubble populations. The probability density of an observed feature vector $\tilde{\mathbf{x}}$ under the mixture model is given by
\begin{equation}
p(\tilde{\mathbf{x}}) =
\sum_{j=1}^{2} \pi_j \,
\mathcal{N}\!\left(
\tilde{\mathbf{x}} \mid \boldsymbol{\mu}_j, \boldsymbol{\Sigma}_j
\right),
\label{eq:gmm}
\end{equation}
where $\pi_j$ denotes the mixing coefficient of the $j$th Gaussian component, $\boldsymbol{\mu}_j$ is the corresponding mean feature vector, and $\boldsymbol{\Sigma}_j$ is the associated covariance matrix. Model parameters were estimated using the expectation–maximization algorithm by maximizing the log-likelihood over all bubble instances within a frame.

For each bubble instance $k$, cluster membership was determined via maximum \emph{a posteriori} (MAP) assignment,
\begin{equation}
z_k = \arg\max_{j} \; p\!\left(j \mid \tilde{\mathbf{x}}_k \right),
\label{eq:map}
\end{equation}
where $z_k \in \{1,2\}$ denotes the assigned cluster index and $p(j \mid \tilde{\mathbf{x}}_k)$ is the posterior probability inferred from the fitted mixture model.

To identify the in-focus cluster in a physically informed manner, the component exhibiting the largest contrast between sharpness and depth-related features was selected. Specifically, the in-focus cluster index $j^\ast$ was defined as
\begin{equation}
j^\ast = \arg\max_{j} \left( \mu_{j,F} - \mu_{j,Z} \right),
\label{eq:incluster}
\end{equation}
where $\mu_{j,F}$ and $\mu_{j,Z}$ denote the mean sharpness and depth-proxy components of $\boldsymbol{\mu}_j$, respectively. This criterion reflects the physical expectation that in-focus bubbles exhibit sharper rims and smaller effective depth proxies than out-of-focus bubbles.

The final binary focus label for bubble $k$ was then assigned as
\begin{equation}
y_k = \mathbb{1}[\, z_k = j^\ast \,],
\label{eq:label}
\end{equation}
where $\mathbb{1}[\cdot]$ is the indicator function. While the GMM provides soft probabilistic separation between focal populations through posterior probabilities, these hard labels enable interpretable visualization and facilitate subsequent temporal stabilization and supervised learning.

\subsection{Temporal Label Stabilization}

Although GMM provides robust framewise separation of focal populations, small variations in appearance across consecutive frames, arising from noise, illumination fluctuations, and segmentation uncertainty, can lead to intermittent label switching. To mitigate this effect without introducing explicit tracking, a lightweight temporal stabilization strategy is applied that exploits short-time coherence in high-speed image sequences.

For each bubble detected at time $t$, a nearest-neighbor association is established in centroid space with bubbles detected in the preceding frame $t-1$. Let $\mathbf{c}_k(t)$ denote the centroid of bubble $k$ at time $t$. A correspondence is accepted only if the Euclidean distance to the closest centroid in the previous frame satisfies
\begin{equation}
\left\lVert \mathbf{c}_k(t) - \mathbf{c}_m(t-1) \right\rVert_2 \le d_{\max},
\label{eq:nn}
\end{equation}
where $d_{\max}$ is a prescribed spatial threshold that enforces local spatiotemporal continuity. This association does not assume persistent bubble identity, but instead leverages the short-time coherence inherent to densely sampled image sequences.

Once a correspondence is established, temporal consistency is enforced using a focus-based hysteresis rule. If the framewise GMM assignment at time $t$ differs from the stabilized label at time $t-1$, the update is accepted only when the current focus score $F_k(t)$ exceeds a percentile-based threshold $\tau$ derived from the focus-score distribution of the current frame. The stabilized label $y_k(t)$ is thus defined as
\begin{equation}
y_k(t) =
\begin{cases}
1, & y_m(t-1) = 1 \;\text{and}\; F_k(t) \ge \tau, \\
0, & y_m(t-1) = 0 \;\text{and}\; F_k(t) < \tau, \\
z_k(t), & \text{otherwise},
\end{cases}
\label{eq:hysteresis}
\end{equation}
where $y_m(t-1)$ denotes the stabilized label of the associated bubble in the previous frame and $z_k(t)$ is the instantaneous GMM label at time $t$.

This asymmetric update rule suppresses spurious label oscillations while permitting genuine transitions in focal state. Importantly, the stabilization operates entirely on physically interpretable quantities and does not alter the underlying clustering. 
% Instead, it regularizes the temporal evolution of labels in a manner consistent with the gradual nature of optical defocus, thereby improving robustness without imposing hard temporal constraints or explicit motion models.

\subsection{Results Across Void Fractions}

Figure~\ref{fig:low-vf-clustering} presents representative results of the unsupervised clustering pipeline applied to low void-fraction frames, where bubble overlap is limited and individual interfaces are well resolved. In this regime, the GMM-based separation of in-focus and out-of-focus bubbles is stable and visually consistent across consecutive frames, with temporal stabilization effectively suppressing spurious label switching. The resulting clustered outputs exhibit clear focal stratification and serve as reliable depth-aware representations of the dispersed phase under dilute conditions.

Figure~\ref{fig:high-vf-clustering} shows the corresponding clustering results for high void-fraction frames, where frequent overlap, and partial occlusion introduce greater ambiguity in feature distributions. While the pipeline continues to produce physically interpretable groupings, the separation between focal populations becomes increasingly sensitive to void fraction, with clustering boundaries shifting in response to local congestion and altered edge statistics. These observations indicate that, although effective within individual regimes, the unsupervised GMM formulation is not inherently invariant to void fraction.

As hypothesized at the outset, the dependence of feature distributions on void fraction motivates adaptive treatment across flow regimes. Rather than enforcing void-fraction invariance within the unsupervised model itself, the clustered outputs obtained from both low and high void-fraction conditions are instead leveraged as pseudo-labels for supervised learning. By aggregating these regime-specific unsupervised predictions into a unified training set, the subsequent supervised model is exposed to a broader range of bubble appearances and interaction states, enabling it to learn decision boundaries that generalize across void fractions.

% The following section introduces a supervised learning framework trained on these combined pseudo-labeled datasets, with the objective of achieving a void-fraction-invariant bubble classification and tracking pipeline.

%%%%%%%%%%%%%%%%%%%%%%%%%%%%%%%%%%%%%%%%%%%%%%%%%%%%%%%%%%%

%% Use title case for subsections and subsubsections (first letter of words capitalized)
\section{METHODS \& RESULTS - SUPERVISED LEARNING}

\subsection{Overview}

To address the void-fraction sensitivity observed in unsupervised clustering, a supervised learning model was trained using pseudo-labels generated across multiple flow regimes. 
% In contrast to the unsupervised stage, which provides relative focal stratification, the supervised component targets reliable detection of in-plane bubbles and aims to generalize across variations in bubble density and image appearance.

% A pixel-wise segmentation model was developed to detect in-plane bubbles in high-speed image sequences. 
The approach employs a Random Forest (RF) classifier \cite{xi2022_rf_dl} operating on a compact set of physically interpretable image features that build on Section~\ref{sec:ULfeature} and extend to generalized structures. This design was guided by a practical constraint: both training and inference needed to remain computationally feasible on standard CPU hardware for large image volumes. Thus, we adopted a lightweight, feature-based supervised model rather than a data-intensive deep learning architecture.

\subsection{Dataset Split and Supervision}

\begin{figure}[h]
\centering
\includegraphics[width=\textwidth,height=0.3\textheight,keepaspectratio]{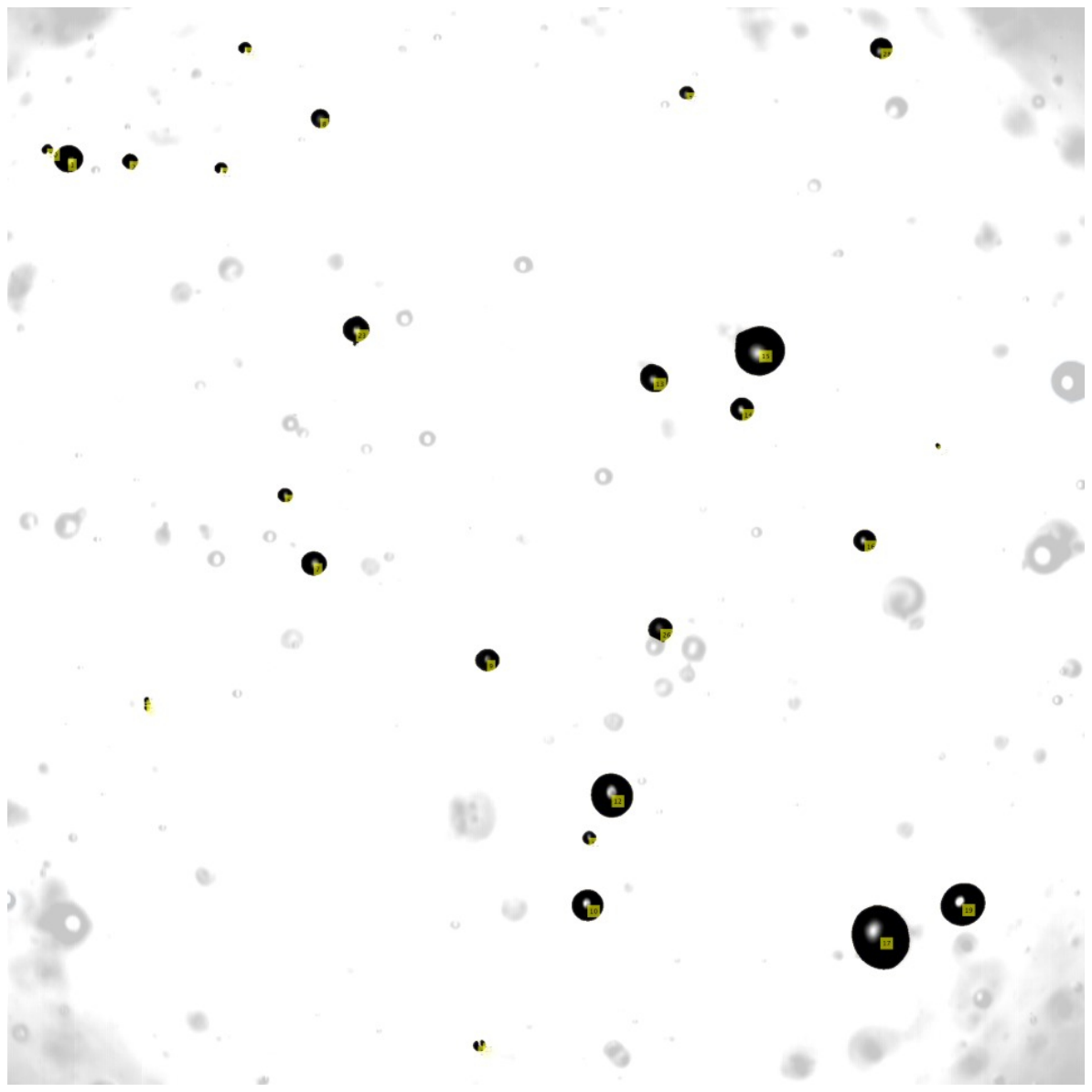}
\caption{Representative frame employed as Ground Truths for the Random Forest supervised model, generated from \cite{kumar2025bubble}.}
\label{fig:groundtruth}
\end{figure}

Limited ground-truth labels were generated using the canny edge-detection method and morphological changes to identify in-plane bubbles as shown in Figure~\ref{fig:groundtruth}. This, alongside pseudo-labels generated for both in-plane and out-of-plane bubbles using GMM were used as the whole labeled dataset for the RF classifier. To mitigate correlation effects between successive frames, model development employed a frame-wise data split.

\subsection{Pre-processing}

Pre-processing was applied prior to feature extraction to ensure consistency across frames and to reduce computational overhead. First, frame-to-frame illumination bias was suppressed so that intensity-based features remained comparable across the dataset. Second, spatial downscaling was applied to reduce computational cost while preserving the dominant bubble-scale structures relevant for classification. These steps ensured that the supervised model operated on standardized inputs while maintaining efficiency suitable for large, high-speed image sequences.

After the illumination correction performed in Section~\ref{sec:illuminationcorrection}, as a layer of insurance, background subtraction was performed to negate spurious effects, if any were left after normalized correction. 
% Each frame is first converted to grayscale and normalized to unit intensity to ensure consistent dynamic range across the dataset. To suppress large-scale illumination gradients and background shading artifacts, a low-frequency background component is estimated using Gaussian smoothing and subsequently subtracted from the original image. This operation enhances local contrast associated with bubble interiors and boundaries while reducing spurious responses arising from nonuniform lighting.

Specifically, the illumination-corrected image $I_c(x,y)$ at pixel location $(x,y)$ was computed as
\begin{equation}
I_c(x,y) = \mathcal{N}\!\left( I(x,y) - \alpha \, \big(G_\sigma * I\big)(x,y) \right),
\label{eq:bg}
\end{equation}
where $I(x,y)$ denotes the normalized intensity, $G_\sigma$ is a Gaussian kernel with standard deviation $\sigma$ controlling the spatial scale of background variation, and $*$ denotes convolution. The scalar coefficient $\alpha \in (0,1)$ modulates the strength of background subtraction, allowing partial suppression of low-frequency intensity components while preserving bubble-scale structure. The operator $\mathcal{N}(\cdot)$ denotes min–max normalization applied to rescale the corrected image to the unit interval.

To reduce computational cost while preserving bubble-scale structure, all images were downscaled by a factor of four in both spatial dimensions prior to feature extraction. Specifically, original frames of size $2048 \times 2048$ pixels were resampled to $512 \times 512$ pixels, resulting in a sixteen-fold reduction in the number of pixels processed. This downscaling substantially improved processing speed while retaining the dominant spatial features required for bubble detection and classification.

\subsection{Feature Engineering}

Following pre-processing, a compact set of pixel-wise features was engineered to capture complementary visual metrics while remaining computationally efficient. For each pixel location $(x,y)$, a six-dimensional feature vector was constructed as shown in Eq.~\ref{eqn:feature_layout}.

\begin{equation}
\begin{aligned}
\mathbf{f}(x,y) = \Big[ &
I(x,y),\;
G_{\mathrm{Sobel}}(x,y),\;
T(x,y), \\
& \mathrm{LoG}_{\max}(x,y),\;
\sigma_{\mathrm{index}}(x,y),\;
\mathrm{RMS}_{\mathrm{local}}(x,y)
\Big]
\end{aligned}
\label{eqn:feature_layout}
\end{equation}

These features provide sensitivity to bubble interfaces, blob-like structures, characteristic spatial scales, and local contrast context, while keeping the representation low-dimensional and suitable for Random Forest classification.

The intensity feature $I(x,y)$ represents the corrected grayscale brightness at each pixel. Although intensity alone is weakly informative due to its sensitivity to experimental conditions and illumination variability, it provides a useful prior when fused with gradient and structure based features.

In-plane bubbles typically exhibit pronounced rim boundaries arising from refractive-index contrast as explained in Section~\ref{sec:ULfeature}. The Sobel gradient magnitude quantifies this edge strength and serves as a robust indicator of bubble boundaries while guarding against false positives driven by intensity alone. Directional gradients are computed along the horizontal and vertical axes as
\begin{equation}
\begin{aligned}
G_x &= \mathrm{Sobel}(I, dx = 1, dy = 0), \\
G_y &= \mathrm{Sobel}(I, dx = 0, dy = 1),
\end{aligned}
\label{eqn:directional_sobel}
\end{equation}
where $I(x,y)$ denotes the corrected image intensity and the Sobel operator approximates first-order spatial derivatives using discrete convolution kernels. The gradient magnitude is then computed as
\begin{equation}
G(x,y) = \sqrt{G_x^2 + G_y^2},
\label{eqn:sobel}
\end{equation}
providing a rotationally invariant measure of local edge strength. This feature highlights sharp bubble rims while suppressing slowly varying background structures and illumination artifacts.

The Tenengrad feature measures local gradient energy and is sensitive to spatially coherent edge structures. In the present context, Tenengrad complements the Sobel magnitude by emphasizing regions where edge sharpness is both strong and spatially consistent, thereby distinguishing between sharp in-plane rims and blurred out-of-plane structures. It is computed as a locally smoothed measure of squared gradient magnitude,
\begin{equation}
T(x,y) = \mathrm{blur}\!\left(G^2(x,y)\right),
\label{eqn:tenengrad}
\end{equation}
where $\mathrm{blur}(\cdot)$ denotes local averaging over a small spatial neighborhood. Unlike the Sobel magnitude, which measures edge strength at a single point, the Tenengrad feature reflects how consistently high-frequency content is present within a local region. This makes it more effective at suppressing blurred structures that might otherwise be misidentified as bubble interfaces.

The Laplacian-of-Gaussian (LoG) maximum response $\mathrm{LoG}_{\max}$, captures second-derivative sensitivity after scale-dependent smoothing and is well suited for bubble detection due to the coherent, closed nature of bubble interfaces. Unlike first-derivative operators such as Sobel, which respond to any linear edge, the LoG operator emphasizes blob-like structures and suppresses elongated edges that do not correspond to bubble geometries, thereby reducing false positives. For a set of smoothing scales $\sigma \in \{1,3,5\}$, the LoG response is computed as
\begin{equation}
\begin{aligned}
I_{\sigma} &= \mathrm{GaussianBlur}(I,\sigma), \\
L_{\sigma} &= \left| \mathrm{Laplacian}(I_{\sigma}) \right|,
\end{aligned}
\label{eqn:logmax_setup}
\end{equation}
The absolute value ensures that both positive and negative second-derivative extrema are treated symmetrically. The maximum LoG response across scales is defined as
\begin{equation}
\mathrm{LoG}_{\max}(x,y) = \max_{\sigma} L_{\sigma}(x,y),
\label{eqn:logmax}
\end{equation}

In addition to the magnitude of the LoG, the scale at which this maximum occurs provides information about the characteristic size of local structures. Because bubbles appear across a range of spatial scales, reliance on a single-scale response is insufficient for robust detection. The LoG scale index is then defined as
\begin{equation}
\sigma_{\mathrm{index}}(x,y) = \arg\max_{\sigma} L_{\sigma}(x,y),
\label{eqn:index_logmax}
\end{equation}
recording the smoothing scale at which the LoG response peaks. This index helps the RF classifier improve its generalization across varying void fractions.

The local root-mean-square (RMS) intensity feature quantifies spatial variability in pixel intensity within a neighborhood surrounding each pixel. By summarizing neighborhood-level variance, the RMS feature captures contrast information that complements gradient and blob-based features.

\subsection{Training and Hyperparameters}
Training was performed on 80\% of the combined ground truth labels and the pseudo-labeled dataset for pixel-wise segmentation. Because pixel-wise labeling produces a highly imbalanced dataset (bubble pixels are sparse relative to background), class balancing was enforced through controlled sampling. All labeled positive pixels (in-plane bubbles) were retained, while negative pixels (background) were randomly subsampled to at most $K=3$ times the number of positives. To further bound memory usage and training time, the total number of training samples was hard-capped at $3\times 10^6$ pixels. This sampling strategy preserved positive examples while preventing background pixels from dominating the learning objective of the supervised model.

Model hyperparameters were selected to provide sufficient representational capacity to separate bubbles from background while limiting variance through ensemble averaging. The RF used 150 trees, a minimum of 5 samples per leaf, and no explicit depth limit (\texttt{max\_depth=None}). Class imbalance was additionally handled through \texttt{class\_weight="balanced\_subsample"}, which reweights classes within each bootstrap sample without requiring a larger training set. Training and inference were parallelized across all available cores (\texttt{n\_jobs=-1}). Together, these settings favor high recall of bubble pixels under varying appearance conditions while keeping computational cost manageable. The minimum leaf size and the averaging effect of multiple trees provide practical regularization despite the unlimited tree depth.

\subsection{Results}

RF showed consistently strong performance on both the training and held-out test sets, indicating that its learned decision rules generalize well to unseen frames under the tested acquisition conditions. Because bubble pixels are a minority class in pixel-wise segmentation, Average Precision is emphasized here as it summarizes the precision-recall tradeoff across thresholds and is not dominated by the large number of background pixels, and the resulting AP was high (0.962 train, 0.818 test).

At the operating threshold, Precision, Recall, and F1 scores for bubble detection were 0.924, 0.776, 0.844 on training and 0.901, 0.488, 0.634 on test, respectively. the False-positive rates remained minimal in both cases (4.4\% train, 6.1\% test). Together, these results indicate that the RF recovers most bubble pixels while maintaining low background contamination (high Precision and low false-positive rate). The train-test gaps across recall suggest the model is conservative, and lenient thresholds will improve the performance further. Overall, the metrics suggest a limited overfitting and provide a quantitative baseline against which subsequent post-processing refinements can be assessed.

\begin{figure}[h]
\centering
\includegraphics[width=\textwidth,height=0.25\textheight,keepaspectratio]{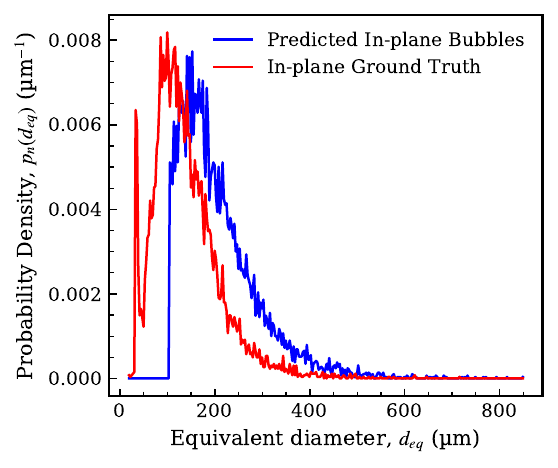}
\caption{Probability density of equivalent diameter $d_{eq}$ for in-plane RF predictions compared with in-plane ground truth, showing a minor upward shift while retaining relatively high agreement in distribution shape (PDF overlap $=0.713$).}

\label{fig:pdf_inplane_pred_gt}
\end{figure}

For the in-plane bubble population, the prediction was compared directly against all available in-plane ground-truth diameters using probability density functions (PDFs). The predicted in-plane set contains $N=10436$ bubbles with $d_{50}=185.66~\mu\mathrm{m}$ and $d_{10}$--$d_{90}=122.61$--$315.24~\mu\mathrm{m}$, while the ground truth contains $N=9418$ bubbles with $d_{50}=125.89~\mu\mathrm{m}$ and $d_{10}$--$d_{90}=64.41$--$224.24~\mu\mathrm{m}$. Small bubbles present in the ground truth are not represented in the predicted in-plane set under the current detection and post-processing settings, possibly due to stricter thresholds. Figure~\ref{fig:pdf_inplane_pred_gt} shows the corresponding PDF comparison.

\begin{figure}[h]
\centering
\includegraphics[width=\textwidth,height=0.25\textheight,keepaspectratio]{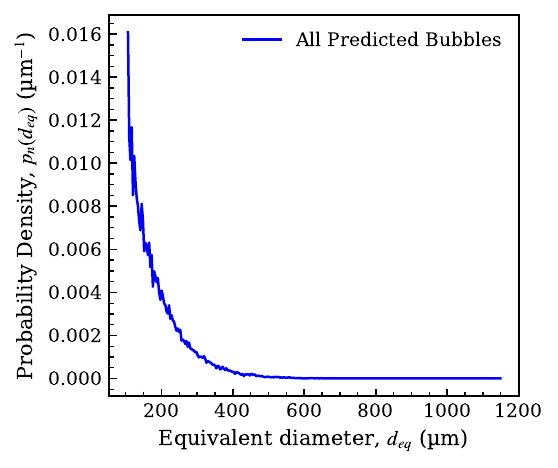}
\caption{Probability density of equivalent diameter $d_{eq}$ for all predicted bubbles, dominated by out-of-plane detections, with a long upper tail and dense minimum $d_{eq}$ sized bubbles.}

\label{fig:pdf_all_pred}
\end{figure}

Across all predicted bubbles, the distribution is dominated by out-of-plane detections. $N=64079$ total bubbles versus $N=10436$ in-plane bubbles, implying that $\approx 16.3\%$ of detected bubbles are classified as in-plane. The all-bubbles PDF yields $d_{50}=161.35~\mu\mathrm{m}$ with $d_{10}$--$d_{90}=112.48$--$290.10~\mu\mathrm{m}$ and a long upper tail reaching $d_{\max}=1150.63~\mu\mathrm{m}$. Because out-of-plane bubbles constitute the majority of detections ($N=53643$), its statistics closely track the aggregate (median $156.74~\mu\mathrm{m}$ versus all-bubbles median $161.35~\mu\mathrm{m}$). Figure~\ref{fig:pdf_all_pred} reports the PDF for the full predicted population and provides the baseline size distribution.

\begin{figure}[h]
\centering
\includegraphics[width=\textwidth,height=0.25\textheight,keepaspectratio]{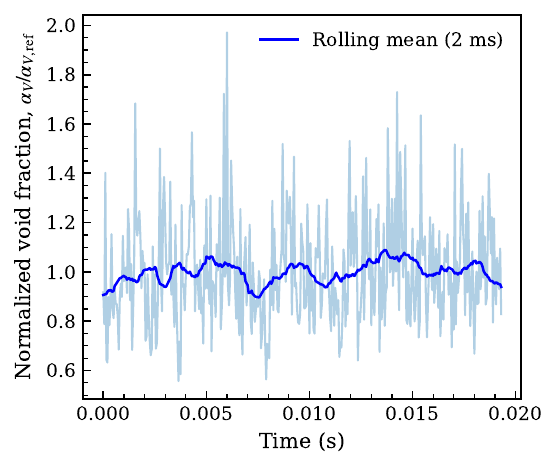}
\caption{Normalized void fraction time series for all predicted bubbles.}
\label{fig:void_fraction_all_pred_norm}
\end{figure}

Across the total labeled frame sequence ($\approx 20~\mathrm{ms}$ at 20000 FPS), a volumetric void fraction estimate was computed from the predicted bubble sizes under a spherical-bubble assumption and a fixed control-volume depth. For each frame, the void fraction was defined as
\begin{equation}
\alpha_V(t)=\frac{\sum_{i\in \text{frame }t} V_i}{V_{\mathrm{CV}}}
\end{equation}
\begin{equation}
V_i=\frac{\pi}{6}d_{eq,i}^{\,3}
\end{equation}
\begin{equation}
V_{\mathrm{CV}}=A_{\mathrm{ROI}}\,H,    
\end{equation}
where $d_{eq,i}$ is the equivalent diameter of bubble $i$, $A_{\mathrm{ROI}}$ is the region-of-interest cross-sectional area, and $H=1.5~\mathrm{mm}$ is the assumed control-volume depth. The void fraction with time series is reported in normalized form, $\alpha_V(t)/\alpha_{V,\mathrm{ref}}$, where $\alpha_{V,\mathrm{ref}}=\langle \alpha_V\rangle$ is the mean over the sequence. For all predicted bubbles, the normalized series had median $0.970$ with a $10$--$90$th percentile range $0.756$--$1.275$, indicating bounded fluctuations about the reference level. Figure~\ref{fig:void_fraction_all_pred_norm} visualizes the normalized void fraction with a 2~ms rolling mean.

% \begin{figure}[h]
% \centering
% \includegraphics[width=\textwidth,height=0.2\textheight,keepaspectratio]{SL_lowvoid.pdf}
% \caption{Post-trained RF classified frame classifying bubbles as in-focus and out-of focus for a relatively lower void fraction.}
% \label{fig:SL_lowvoid}
% \end{figure}

\begin{figure}[h]
\centering
\includegraphics[width=\textwidth,height=0.2\textheight,keepaspectratio]{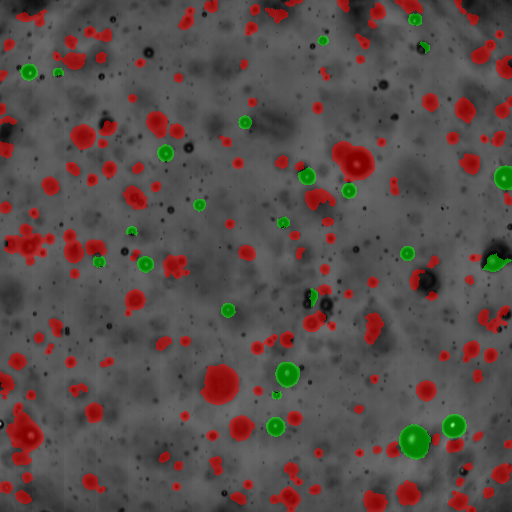}
\caption{Post-trained RF classified frame showing in-plane and out-of-plane bubbles at relatively high void fraction. Non-spherical edge bubbles are retained, and overlapping in-plane bubbles are separated into distinct instances.}

\label{fig:SL_highvoid}
\end{figure}

Figure~\ref{fig:SL_highvoid} show representative predictions with in-plane masks exhibiting consistent boundary placement and minimal spurious detections, qualitatively matching the labeled appearance in Figure~\ref{fig:groundtruth}. The visual agreement is on par with the reported test-set metrics, and suggests that the learned feature-based decision rules remain stable across changes in bubble density and local clutter. The quantitative scores and qualitative overlays indicate that the RF provides a reliable in-plane segmentation baseline for downstream bubble statistics and tracking within the tested acquisition regime.

%%%%% Conclusions %%%%%%%%%%%%%%%%%%%%%%%%%%%%%%%

\section{Conclusion}
Across domains of Science \& Engineering, a ton of imaging and computer vision is applied to an experimental and physical setup. During flow visualisation, particles and masses are tracked to study flow behavior. Tracking requires edge and dimension detection. Image datasets produced can be processed through classical vision algorithms. Although these algorithms do a great job in terms of image characteristics, ML algorithms can be tailored to provide a one-step ahead clustering and classification of these physical quantities, image characteristics used by these classical algorithms, and bin these quantities into different clusters. These clusters can then be further mapped to a physical relevance via a score, as we tried to regenerate a 3D relative understanding from a 2D image. Unsupervised Learning had its' tradeoffs of regime limitation of void fraction, for which it was used with a supervised model to produce robustness across void fraction and to generalize the whole process. This helps us study a 2D frame in an aided 3D visualization and track, segment, cluster, and map features to physical relevance, train, and test in a robust pipeline.

%%%%% Acknowledgments %%%%%%%%%%%%%%%%%%%%%%%%%%%

\section*{Acknowledgments}
The information, data, or work presented herein was funded in part by the Advanced Research Projects Agency-Energy (ARPA-E), U.S. Department of Energy, under Award Number DE-AR0001587. The views and opinions of authors expressed herein do not necessarily state or reflect those of the United States Government or any agency thereof. The authors would like to acknowledge Mr. Grant Breckenridge (M.S. Mechanical Engineering, Georgia Institute of Technology, Fall 2025) for his contributions during the initial stages of this project, including exploratory work on U-Net CNN based implementations. In disclosure, the part of this work was presented at APS-DFD~\citep{suchandra2025surfactant,aps_kumar2024_exp}. 

%%%  REFERENCES  %%%%%%%%%%%%%%%%%%%%%%%%%%%%%%%%
%%
%% Put your references into your .bib file in the usual way. Run latex once, bibtex once, then latex twice.
%% The asmeconf.bst style allows @inproceedings and @proceedings to include: 
%%		venue = {Location of Conference}, 
%%		eventdate = {Month, days},

% \nocite{*}%% <=== Delete this line unless you want to typeset the entire contents of your .bib file !!

\bibliographystyle{asmeconf}  %% .bst file following ASME conference format. Do not change.
\bibliography{bibliography}

@misc{sklearn_outlier_2025,
  author       = {{scikit-learn developers}},
  title        = {Novelty and outlier detection},
  howpublished = {scikit-learn documentation},
  year         = {2025},
  url          = {https://scikit-learn.org/stable/modules/outlier_detection.html},
  note         = {Accessed: Oct. 2, 2025}
}

@article{kumar2025bubble,
  title={Bubble coalescence dynamics in a high-Reynolds number decaying turbulent flow},
  author={Kumar, Vivek and Suchandra, Prasoon and Javadi, Ardalan and Jain, Suhas S and Aidun, Cyrus},
  journal={arXiv preprint arXiv:2509.05888},
  year={2025}
}

@inproceedings{suchandra2025surfactant,
  title={Surfactant-induced suppression of bubble coalescence in turbulent duct flow at high Reynolds numbers},
  author={Suchandra, Prasoon and Kumar, Vivek and Rom, Jason and Prajapati, Shivam and Javadi, Ardalan and Jain, Suhas S and Aidun, Cyrus K},
  booktitle={Division of Fluid Dynamics Annual Meeting 2025},
  year={2025},
  organization={APS}
}

@article{aps_kumar2024_exp,
  title={On the coalescence of bubbles in highly turbulent flows},
  author={Kumar, Vivek and Javadi, Ardalan and Jain, Suhas and Aidun, Cyrus},
  journal={Bulletin of the American Physical Society},
  year={2024},
  publisher={APS}
}

@article{sena2025machine,
  title={A Machine Learning Model for the Prediction of Water Contact Angles on Solid Polymers},
  author={Sena, Jose and Johannissen, Linus O and Blaker, Jonny J and Hay, Sam},
  journal={The Journal of Physical Chemistry B},
  volume={129},
  number={10},
  pages={2739--2745},
  year={2025},
  publisher={ACS Publications}
}

@article{kumar2025viscosity,
  title={Viscosity and dynamic surface tension measurement: A guideline for appropriate measurement},
  author={Kumar, Vivek and Quintero, JSM and Baldygin, Aleksey and Molina, Paul and Willers, Thomas and Waghmare, Prashant R},
  journal={arXiv preprint arXiv:2510.05481},
  year={2025}
}

@article{javadi2026large,
  title={Large eddy simulations of side channel pump in different operating conditions},
  author={Javadi, Ardalan and Kumar, Vivek and Aidun, Cyrus K},
  journal={Engineering Applications of Computational Fluid Mechanics},
  volume={20},
  number={1},
  pages={2587723},
  year={2026},
  publisher={Taylor \& Francis}
}

@article{kumar2023particle,
  title={Particle separation using modified Taylor’s flow},
  author={Kumar, Vivek and Jain, Palak and Upadhyay, Ravi Kant and Bharath, KS and Waghmare, Prashant R},
  journal={Microfluidics and Nanofluidics},
  volume={27},
  number={10},
  pages={66},
  year={2023},
  publisher={Springer}
}

@article{sotak1989_log,
  author  = {Sotak, G. E. and Boyer, K. L.},
  title   = {The Laplacian-of-Gaussian Kernel: A Formal Analysis and Design Procedure for Fast, Accurate Convolution and Full-Frame Output},
  journal = {Computer Vision, Graphics, and Image Processing},
  year    = {1989},
  volume  = {48},
  number  = {2},
  pages   = {147--189},
  month   = nov,
  doi     = {10.1016/S0734-189X(89)80036-2}
}

@article{mackiewicz1993_pca,
  author  = {Ma{\'c}kiewicz, A. and Ratajczak, W.},
  title   = {Principal Components Analysis (PCA)},
  journal = {Computers \& Geosciences},
  year    = {1993},
  volume  = {19},
  number  = {3},
  pages   = {303--342},
  month   = mar,
  doi     = {10.1016/0098-3004(93)90090-R}
}

@article{haas2020_bubcnn,
  author  = {Haas, T. and Schubert, C. and Eickhoff, M. and Pfeifer, H.},
  title   = {BubCNN: Bubble Detection Using Faster R-CNN and Shape Regression Network},
  journal = {Chemical Engineering Science},
  year    = {2020},
  volume  = {216},
  pages   = {115467},
  month   = apr,
  doi     = {10.1016/j.ces.2019.115467}
}

@article{toyama2025_fasterrcnn,
  author  = {Toyama, K. and Kanemoto, R. and Misumi, R. and Araki, T. and Mitsushima, S.},
  title   = {Faster R-CNN-Based Detection and Tracking of Hydrogen and Oxygen Bubbles in Alkaline Water Electrolysis},
  journal = {Electrochemistry},
  year    = {2025},
  volume  = {93},
  number  = {2},
  pages   = {027011},
  month   = feb,
  doi     = {10.5796/electrochemistry.24-00127}
}

@article{kim2021_universal,
  author  = {Kim, Y. and Park, H.},
  title   = {Deep Learning-Based Automated and Universal Bubble Detection and Mask Extraction in Complex Two-Phase Flows},
  journal = {Scientific Reports},
  year    = {2021},
  volume  = {11},
  number  = {1},
  month   = apr,
  doi     = {10.1038/s41598-021-88334-0}
}

@article{kartakoullis2025_gmm,
  author  = {Kartakoullis, A. and Caporaso, N. and Whitworth, M. B. and Fisk, I. D.},
  title   = {Gaussian Mixture Model Clustering Allows Accurate Semantic Image Segmentation of Wheat Kernels from Near-Infrared Hyperspectral Images},
  journal = {Chemometrics and Intelligent Laboratory Systems},
  year    = {2025},
  volume  = {259},
  pages   = {105341},
  month   = apr,
  doi     = {10.1016/j.chemolab.2025.105341}
}

@article{xi2022_rf_dl,
  author  = {Xi, E.},
  title   = {Image Classification and Recognition Based on Deep Learning and Random Forest Algorithm},
  journal = {Wireless Communications and Mobile Computing},
  year    = {2022},
  volume  = {2022},
  pages   = {2013181},
  month   = jun,
  doi     = {10.1155/2022/2013181}
}

@misc{gatech_rbi_multiphase_2025,
  author       = {Martin, J.},
  title        = {Georgia Tech’s {RBI} Unveils State-of-the-Art Multiphase Forming Lab},
  howpublished = {Georgia Institute of Technology Research (news/feature)},
  year         = {2025},
  url          = {https://research.gatech.edu/feature/multiphase-forming-lab},
  note         = {Accessed: Oct. 3, 2025}
}

@misc{sklearn_gmm_selection_2025,
  author       = {{scikit-learn developers}},
  title        = {Gaussian mixture model selection},
  howpublished = {scikit-learn example gallery},
  year         = {2025},
  url          = {https://scikit-learn.org/stable/auto_examples/mixture/plot_gmm_selection.html},
  note         = {Accessed: Nov. 7, 2025}
}

@misc{azad2025_losses_segmentation,
  author        = {Azad, Reza and Heidary, Moein and others},
  title         = {Loss Functions in the Era of Semantic Segmentation: A Survey and Outlook},
  howpublished  = {arXiv preprint},
  year          = {2025},
  month         = dec,
  eprint        = {2312.05391},
  archivePrefix = {arXiv}
}

@misc{gfg_unet_2025,
  author       = {{GeeksforGeeks contributors}},
  title        = {U-Net Architecture Explained},
  howpublished = {GeeksforGeeks},
  year         = {2025},
  url          = {https://www.geeksforgeeks.org/machine-learning/u-net-architecture-explained/},
  note         = {Accessed: Dec. 2, 2025}
}

@article{sinha2025_maskrcnn_unet,
  author  = {Sinha, M. and Paul, S. and Nee Lala, M. G.},
  title   = {Comparative Analysis of Mask R-CNN and U-Net Architectures Using ResNet as Backbone for Lunar Crater Detection},
  journal = {Planetary and Space Science},
  year    = {2025},
  volume  = {264},
  pages   = {106140},
  month   = sep,
  doi     = {10.1016/j.pss.2025.106140}
}

@article{Otsu1979,
  author  = {Otsu, Nobuyuki},
  title   = {A Threshold Selection Method from Gray-Level Histograms},
  journal = {IEEE Transactions on Systems, Man, and Cybernetics},
  volume  = {9},
  number  = {1},
  pages   = {62--66},
  year    = {1979},
  doi     = {10.1109/TSMC.1979.4310076}
}

@article{Choi2022BubbleVelocimetry,
  author    = {Daehyun Choi and Hyunseok Kim and Hyungmin Park},
  title     = {Bubble velocimetry using the conventional and CNN-based optical flow algorithms},
  journal   = {Scientific Reports},
  volume    = {12},
  pages     = {11879},
  year      = {2022},
  doi       = {10.1038/s41598-022-16145-y}
}

@article{Hessenkemper2024_3DBubbleTracking,
  author  = {Hessenkemper, Hendrik and Wang, Lantian and Lucas, Dirk and Tan, Shiyong and Ni, Rui and Ma, Tian},
  title   = {3D detection and tracking of deformable bubbles in swarms with the aid of deep learning models},
  journal = {International Journal of Multiphase Flow},
  volume  = {179},
  pages   = {104932},
  year    = {2024},
  doi     = {10.1016/j.ijmultiphaseflow.2024.104932}
}

%%%%%%%%%%%%%%%%%%%%%%%%%%%%%%%%%%%%%%%%%%%%%%%%%%%%%%%%%%%%%%%%%%%%%%%%%%%%%%%%%%%%%%%

\end{document}